\documentclass[11pt,twocolumn]{article}
\usepackage{fullpage}
\usepackage[round]{natbib}
\usepackage{multirow}
\usepackage{amsmath,amsfonts,amssymb,amsthm,  mathrsfs,mathtools}

\usepackage[hyphens]{url}
\usepackage[hyperindex,breaklinks]{hyperref}
\usepackage{tikz,lipsum,lmodern}
\usepackage[most]{tcolorbox}
\usepackage{xcolor}
\usepackage{booktabs}

\hypersetup{
  colorlinks,
  citecolor=violet,
  linkcolor=blue,
  urlcolor=blue}
\usepackage{dsfont}
\numberwithin{equation}{section}   
\usepackage{booktabs} % For formal tables
\usepackage[ruled,noend]{algorithm2e} % For algorithms

\SetAlFnt{\small}
\SetAlCapFnt{\small}
\SetAlCapNameFnt{\small}
\SetAlCapHSkip{0pt}
\IncMargin{-\parindent}

\numberwithin{equation}{section}
\usepackage{macros-common}
\usepackage{macros-project-specific}
\usepackage{slivkins-theorems}

\usepackage[suppress]{color-edits}

\addauthor{ss}{magenta} % for Suho
\addauthor{kr}{purple} % for Keivan
\addauthor{sf}{red} % for Soheil

\title{Online Advertisements with LLMs: Opportunities and Challenges
\footnote{Alphabetical author ordering.}
}

  \newcommand{\country}[1]{#1.}
  \newcommand{\city}[1]{#1}
  \newcommand{\institution}[1]{#1}
  \newcommand{\email}[1]{Email: \texttt{#1}}
  \newcommand{\affiliation}{\thanks}

\author{
 {Soheil Feizi
 \affiliation{
   \institution{University of Maryland}
   \city{College Park, MD}
   \country{USA}
 \email{(sfeizi,hajiagha,krezaei,suhoshin)@umd.edu}
 }}
 \and
 MohammadTaghi Hajiaghayi\footnotemark[2]
 \and
 Keivan Rezaei\footnotemark[2]
 \and
 Suho Shin\footnotemark[2]
}

\date{}
\begin{document}

\maketitle
% Abstract. Note that this must come before \maketitle.
\begin{abstract}   
    % This paper discusses challenges and desiderata to operate an online advertisement system with Large Language Model (LLM).
    % % \slcomment{Although working out the details is part of the point of the paper, it's still too vague to just say ``operating an online ad system with an LLM'' here.}
    % We propose several methodologies to meet those requirements, and verify their validity and discuss advantages and drawbacks of each approach through arguments based on thought experiment.
    % We discuss further technical challenges in implementing and generalizing the system to adapt to practical constraints \sfcomment{We need to add more details to the abstract}
    This paper explores the potential for leveraging Large Language Models (LLM) in the realm of online advertising systems. 
    We introduce a general framework for LLM advertisement, consisting of modification, bidding, prediction, and auction modules. 
    Different design considerations for each module are presented.
    These design choices are evaluated and discussed based on essential desiderata required to maintain a sustainable system.
    Further fundamental questions regarding practicality, efficiency, and implementation challenges are raised for future research.
    Finally, we exposit how recent approaches on mechanism design for LLM can be framed in our unified perspective.
    % Finally, we explore the prospect of LLM-based dynamic creative optimization as a means to significantly enhance the appeal of advertisements to users and discuss its additional challenges.
    % \sscomment{Change abstract}
    % \krcomment{need to rewrite to talk a little bit about personalized ads.}
\end{abstract}

\section{Introduction}

% \kredit{need to rewrite to talk a little bit about personalized ads.}
In the vast landscape of online search engines, the role of advertisements has become pivotal, shaping the digital experience for users globally.
The enormity of the market, with trillions of dollars at play, shows the economic significance of advertising, \eg the market size of search advertisement as of $2020$ was valued at USD $164.12$ billion~\citep{AdScale}.

Online advertising not only brings revenue to platform companies but also plays a crucial role in subsidizing the free access to information and services to every individual.
% In a world where people increasingly expect free access to information and services, advertisements play a crucial role in subsidizing these offerings. 
The democratization of education, information, and other resources is made possible through ad-driven models, enabling individuals to not only consume content for free but also contribute to the digital ecosystem. 
Additionally, the symbiotic relationship between advertisements and content creation creates a feedback loop, fostering economic growth.
As a side note, even subscription-based streaming platforms like Netflix are starting ad-supported plans~\citep{Netflix}.
% Several online platforms often offer a subscription model for people to disable the advertisement in using their services, however, the advertisement market is becoming increasingly dominant, \eg Netflix recently started an ad-supported plan~\cite{Netflix}. 

On the other hand, recently, large language models (LLMs)~\citep{brown2020language,anil2023palm,thoppilan2022lamda} have gained widespread adoption among users,
serving various functions including question answering, content generation, translation, code completion,
and more \citep{nijkamp2022codegen, fried2022incoder, wang2021gpt, liu2023visual}.
The proliferation of AI-driven assistant language models, such as ChatGPT,
has contributed to a growing trend wherein individuals increasingly use such models to address their inquiries,
occasionally replacing traditional search engines as their primary information-seeking tool.
According to~\cite{PCMag}, even for now, $35\%$ of casual users say they find LLMs to be more helpful in finding information than search engines.
It is obvious that such a trend will be accelerating in the near future as well.
The substantial usage volumes stemming from diverse users would induce companies offering these tools, which we call \textit{LLM providers},
to contemplate revenue generation through advertising~\citep{Adweek, Techcrunch, Microsoft}.
Consequently, an interesting and fundamental question arises: 

\definecolor{mycolor}{rgb}{0.9, 0.90, 0.9}
\begin{tcolorbox}[colback=mycolor,colframe=gray!75!black]
\begin{center}
  How can \textbf{LLM providers} make revenue by running an online \textbf{advertisement} on their services?
  \end{center}
\end{tcolorbox}

The concept of online advertising has been extensively studied within the realm of search engines, where auctions are conducted among advertisements from advertisers when a user inputs a query.
This paper focuses on the prospect of transposing this online advertising model and auction framework to the context of large language models.
We further discuss technical challenges and potential framework to run online advertisement system in LLM, thereby calling academic and industrial researchers to the area of importance.

\begin{figure*}
    \begin{center}
    \includegraphics[width=\linewidth]{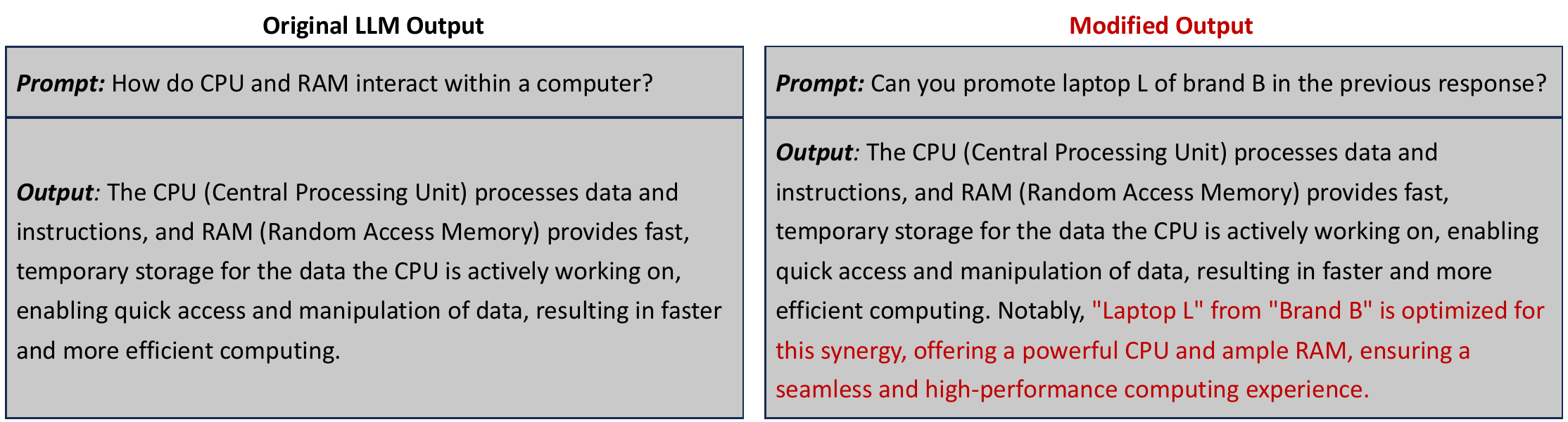}
    \end{center}
    \vspace{-2mm}
    \caption{An example of providing \textbf{unstructured} advertisement in the LLM. Left column refers to the case where we ask above question from ChatGPT 4,
    and to \textbf{incorporate the ads} we use queries specified in the right column.}
    \label{fig:intro}
\end{figure*}

 % In the \textbf{right panel}, the user is interested in machine learning conferences in Florida.
        % Chat bot is advertising ``SunShine Travel" which is a fictional travel agency by putting its logo on a general image from Florida (left)
        % Incorporating personalized advertisement, model can generate more specific advertisements by offering flight tickets
        % from the place that the users interacts with the system and showing a more related image of Florida.
        % For example, when a young unmarried person from DC interacts with the system,
        % output image could advertise tickets from DC to Florida and could be modified to more appealing for the user (middle)
        % or when a married person with their family living in California enters the query, the image could promote flight tickets from there to Florida
        % while showing an image of amusement parks in destination.

\paragraph{Search advertising}
To better explain fundamental differences between the standard search advertising (SA)\footnote{Its mechanism design problem is often called as sponsored search auction (SSA).} and the LLM advertising (LLMA), we briefly introduce how the standard SA works~\citep{lahaie2007sponsored} in what follows.
(1) {\em Bidding}: in the SA, the owner of each ad $i$ writes bid $b_i \in \R_{\ge 0}$ on targeting \emph{keyword} for $i \in [n]$, which can possibly a set of keywords.
(2) {\em Output generation}:  the platform first decides \emph{how many slots} to allocate for ads in the search engine results page (SERP), say $k$.
(3) {\em Prediction}: given $k$ slots in SERP, the platform then predicts the click-through-rate (CTR) $\alpha_{ij}$  when ad $i$ is allocated in slot $j$.
(4) {\em Auction}: then it optimizes 
\begin{align}
    \max_{x \in [0,1]^{n \times k}}\sum_{i=1}^n \sum_{j=1}^k \alpha_{ij}b_i x_{ij},\label{eq:alloc-sa}
\end{align}
where $x = (x_{ij})_{i \in [n], j \in [k]}$ is the (possibly randomized) allocation vector such that $x_{ij} = 1$ if ad $i$ is allocated in slot $j$ given the constraint $\sum_{i=1}^n x_{ij} \le 1$ for every $j \in [k], i \in [n]$, and charges each ad by the committed payment rule.

Overall, whenever a user arrives in the platform and searches a keyword, the set of ads related to the keyword are getting involved in the mechanism.
Specifically, the platform collects corresponding bids, decides the number of slots $k$, predicts CTR, and the auction runs.

% \begin{enumerate}
%     \item Bidding: In the SSA, the owner of each ad $i$ writes bid $b_i \in \R_{\ge 0}$ on targeting \emph{keyword} for $i \in [n]$, which can possibly a set of keywords.
%     \item User interaction: Whenever a user arrives and searches a keyword, the advertisers who specified that keyword automatically participates in an auction.
%     \item Output generation: The platform first decides \emph{how many slots} to allocate for ads in the search engine results page (SERP), say $k$.
%     \item Prediction: Given $k$ slots in SERP, the platform then predicts the click-through-rate (CTR) $\alpha_{ij}$  when ad $i$ is allocated in slot $j$.
%     \item Auction: the mechanism optimizes $\sum_{i=1}^n \sum_{j=1}^k \alpha_{ij}b_i x_{ij}$, where $x_{ij} = 1$ if ad $i$ is allocated in slot $j$, given the constraint $\sum_{i=1}^n x_{ij} \le 1$ for every $j \in [k]$, and charges each ad with respect to precommitted payment rule.
% \end{enumerate}

% If the platform runs a generalized second price auction, which is fairly common in practice~\cite{milgrom2004putting}, it sorts the ads by $\alpha_1 b_1 \ge \alpha_2 b_2 \ge \ldots \ge \alpha_n b_n $, and select top $k$ ads.
% Each selected ad is charged the lowest bid necessary to retain its position, \ie $\ell$-th highest selected ad is charged $\alpha_{\ell + 1}b_{\ell+1}/ \alpha_{\ell}$ for $\ell \in [k]$.

\paragraph{Motivating example}
How would the LLMA be fundamentally different from the SA?
We start with illustrative scenarios where a user asks a technical question about computers (Figure~\ref{fig:intro}).
Without advertisement, LLM would typically generate a response to address the user's query.
% To incorporate advertisements in the generated response and induce user attention, there is a spectrum of possibilities for including ad content: (a)  putting the ads outside the response but in the user interface, (b) incorporate the ads within the generated output directly. \krcomment{later we refer to structured and unstructured output (e.g. Figure~1 caption), but they are not defined here -- is footnote enough?}
% \sscomment{I thought the contents in footnote a bit hurt the flow here, and it might be fine to just let it be in footnote. Do you think we should move it to the main body?}
(a) can be deemed as display ads, and would be relatively easy to handle given the vast amount of techniques in display ads.
However, (b) is more similar to search ad or native ad.\footnote{Search/native ads usually capture user attention better than display ads, \eg almost $50\%$ more views~\citep{NativeSearch}.}
We will focus on the approach (b), which will entail fundamental challenges that have \emph{not arisen} in the traditional SA.
\footnote{Also, in case the ads are included within the generated output, there are several possibilities such that the ads may replace one of the element in a list of elements (\eg Figure~\ref{fig:intro2} in the appendix, denoted by \emph{structured} output), or the ads itself is incorporated within the original text/figure without hurting/changing the context of the original output (\eg Figure~\ref{fig:intro}, denoted by \emph{unstructured} output). We focus on the unstructured output, as this would be broadly applicable in various scenarios, and the structured output may be able to handled with the standard SA framework.}
% (a) ads alongside the output, (b) explicitly putting advertisements in the output like display ads, and (c) including advertisements within the generated output.

\paragraph{SA versus LLMA}
Recall the process of bidding, output generation, prediction, and auction in the SA mentioned before, and imagine implementing those modules for LLMA.

For the bidding module, how could the advertisers \emph{write bid}?
What would they actually \emph{bid for}?
This is not straightforward to answer, since the user query in LLM cannot be simply specified as a keyword.
Further, given that the marketing impact will be significantly dependent to how the LLM incorporate the ad in the output, it is not even clear what is the advertiser's \emph{value} for being included in the ad. 
% \krcomment{I'd merge this and next paragraph (remove the empty line)}
In fact, the advertiser's value would be dependent to the generated output, then again, how could the advertiser reflect their \emph{willingness-to-pay} with respect to the output, which might not be accessible in advance?
Even further, how can we generate output that smoothly incorporates the ad without hurting the user experience while satisfying the advertiser?

For prediction module, most SA run an online learning algorithm to update the ad's feature vector with respect to the user context~\citep{mcmahan2013ad}.
This was possible because the ad images, hyperlinks, and more generally how they appear in the SERP remain the same across many user interactions.
LLM, however, would incorporate ad in a very different manner for each query, \eg see Figure~\ref{fig:personalized-ad}, then how can the LLMA \emph{learn the CTR} fundamentally?
Also, since the ads are merged into the generated output, it significantly affects the user experience, then how can we guarantee the user satisfaction and measure them?

Finally, which kind of auction format should LLMA run? How can LLMA adapt for advertising multiple ads in a single output?
What would be a reasonable analogue of the autobidding system prevalent in the modern online ad system?

All these questions are not straightforward to answer, yet, have not been formally discussed in any literature to the extent of our knowledge.

% \sscomment{Mention more about difference}
% How to construct the output/ how bidders bid/ how to predict CTR/ 

% We will focus on the case where the ads are incorporated in the unstructured output of the LLM, since the other cases could be handled using techniques from modern online search/display advertisement system.
% Moreover, this kind of advertisement can be more interesting for both users and advertisers due to the huge capabilities of language models in generating high-quality responses while promoting products and services from advertisers.\footnote{Indeed, it's worth noting that Microsoft Bing has been running such types of advertisements in its AI chatbot~\cite{Microsoft}.}

% Then, a natural question is how the system can incorporate advertisements into the generated output to satisfy both the advertisers and the user.
% Further, it is not trivial how the advertisers should communicate with the system to bid on each modified output.

\paragraph{Our contribution}
In this paper, we introduce a vast amount of research questions to operate LLMA in practice.
% We first discuss several desiderata to run LLMA and propose a seemingly plausible framework to meet the requirements.
Similar to SA, our framework consists of four modules, though the implementation of each module will be very different from SA:
(i) \emph{modification} in which the original output of LLM is modified;
(ii) \emph{bidding} that advertisers utilize to bid on the modified outputs;
(iii) \emph{prediction} in which LLMA computes required information about advertisements; and
(iv) \emph{auction} in which the advertises compete and the final output is selected.
We introduce several design choices available for each module, which are evaluated based on a reasonable set of criteria required to maintain a sustainable online platform.
We further discuss specific research challenges inherent to each module, and finally propose a unified perspective in how on could interpret recent approaches for LLM advertisement platform within our framework.
Although we believe that our framework may be grounded as an initial stepping stone for future research, our primary purpose is to discuss research questions that should be addressed for the practical operation of LLM advertisement.\footnote{As an independent interest, in Appendix~\ref{sec:personalized-ad},
% as an LLM's (more generally generative model's) use case of independent interest
we discuss how LLM (more generally generative model) can empower the online advertisement system to tailor the content of ads itself to individual users,
leveraging their specific contexts to generate more appealing content, which is called as \emph{dynamic creative optimization} (DCO) or \emph{responsive ad} in the literature.
The incorporation of DCO into the SA, especially for LLMA, would bring further challenges in each module and we propose corresponding research questions therein.}

% boost the modern \emph{dynamic creative optimization} technique to improve both the user and advertiser experience.
% It is worth noting that the substantial capabilities of generative models empower us to tailor the content of ads itself to individual users,
% leveraging their specific contexts to generate more appealing content, which is called as dynamic creative optimization (DCO) or responsive ad in the literature.
% This approach holds the potential to significantly boost revenues for both the platform and advertisers.
% This concept is detailed in Section~\ref{sec:personalized-ad}.
% \krcomment{ensure that intro is done within first two pages. we can shirnk some paragraphs and remove redundant phrases.}

\section{Related works}
 % \krcomment{i'll add some stuff later.}
Here, we discuss related works on LLMs, online ads, and their intersection. Further discussion can be found in Appendix~\ref{app:fur-rel}.

% The intersection of artificial intelligence, natural language processing, and conversational agents has witnessed significant advancements in recent years.
% The advent of the Transformer architecture, introduced in \cite{vaswani2017attention}, laid the foundation for large language models.
% Transformers, with their self-attention mechanism, revolutionized the way models capture contextual information, enabling superior performance on NLP tasks.
% By scaling up those architectures, OpenAI's GPT-3 \citep{brown2020language} and Google's BERT \citep{devlin2019bert} are obtained, revolutionizing the field of chatbots, enabling more natural and contextually relevant interactions. 
% Those models showed remarkable performance on the natural language generation task leading to the development of chatbots and conversational AI systems.
% These chatbots engage in context-aware, human-like conversations, offering valuable solutions in customer service, healthcare, and educational domains \cite{abd2020effectiveness, nicolescu2022human}.

\paragraph{Large Language Models}
Advancements in AI, NLP, and conversational agents, driven by Transformer architecture \citep{vaswani2017attention},
have given rise to models like GPT-3 \citep{brown2020language} and BERT \citep{devlin2019bert}.
These models revolutionize chatbots, enabling context-aware, human-like interactions across diverse domains \citep{abd2020effectiveness, nicolescu2022human}.
Everyday use of language models has led researchers to investigate the content generated by these models to ensure that they do not hallucinate \citep{guerreiro2023hallucinations, Ji_2023, li2023halueval, zhang2023siren} in their outputs, and do not generate harmful or biased content \citep{pmlr-v139-liang21a, navigli2023biases, NEURIPS2021_1531beb7, shen2023large, weidinger2021ethical, liu2023aligning}. In fact, trustworthy of LLMs is actively studied by researchers \citep{liu2024trustworthy}.

% \sscomment{Keivan: Discuss briefly about LLM alignment as we will discuss hallucination sort of things in DCO part.} \krcomment{is that fine?}

\paragraph{Online Advertisement}
Online advertising, particularly within the context of sponsored search auctions, has evolved in recent years,
with notable contributions from prior research.
Sponsored search auctions have been a subject of extensive investigation, emphasizing the optimization of bidding strategies and keyword relevance.\footnote{Online advertisement can be categorized by either of search or display ads. We here focus on search ads.}
\citet{edelman2007internet} provided valuable insights into the economics of sponsored search auctions,
shedding light on the dynamics of keyword auctions.
\citet{goel2009contract} proposed a contract auction between the advertiser and the publisher, and introduce impression-plus-click pricing for sponsored search auction as an application. We refer to the book by \citet{roughgarden2010algorithmic} for more details.
% \slcomment{This citation is a book chapter, not an entire book.}

% \sscomment{Cite Contract Auctions for Sponsored Search}

\paragraph{Mechanism design with LLM agents}
\citet{duetting2023mechanism} recently proposes a mechanism design problem when the bidders are LLM agents.
More specifically, the bidders bid on desired distributions of the \emph{next token}, given the history of the tokens.
They denote this problem by \emph{token auction}, and provide theoretical foundations of what a reasonable mechanism should look like.
Indeed, their model can be considered as a version of our advertiser modification and dynamic bidding framework.
Notably, they require the advertisers to provide their desired distribution and corresponding bidding for every token, which may require significant communication burden to run the mechanism in practice.
In contrast, the advertisers in our framework only bid once for the modified output, thereby inducing less communication cost.
% \footnote{In Appendix~\ref{sec:discuss-comparison}, we provide further discussion on simultaneous works including~\cite{duetting2023mechanism} to compare it with our framework and how one can practically adapt several mechanism design perspectives in online advertisement system within our framework.}

% \paragraph{Auctions for LLM Summary}
% A recent work by 

% \ssedit{
%     Recently, 
% }
% \sscomment{Discuss recent Google paper as well}

% To the extent of our knowledge, however, this is the first paper to discuss technical challenges and feasible architectures in leveraging online advertisement system with LLM.

% Recently, several tech companies started to run Chatbot advertisements.

% \begin{enumerate}
%     \item Chatbot advertising effectiveness: When does the message get through?. This paper is about Facebook sending "sponsored message" to the user, which is very different from what we are doing. This is basically referred to as Chatbot Marketing
% \end{enumerate}
% Several reference links that consider similar questions here, mostly by Microsoft Bing.
% \begin{enumerate}
%     \item \href{https://www.adweek.com/media/microsoft-details-how-advertising-works-on-bings-ai-driven-chat-based-search/}{article 1}
%     \item \href{https://about.ads.microsoft.com/en-us/blog/post/may-2023/a-new-solution-to-monetize-ai-powered-chat-experiences}{article 2}
%     \item \href{https://techcrunch.com/2023/03/29/that-was-fast-microsoft-slips-ads-into-ai-powered-bing-chat/}{article 3}
%     \item \href{https://www.thedrum.com/news/2023/03/30/gpt-4-powered-bing-chat-introduces-ads-and-experts-believe-it-s-natural-fit}{article 4}
% \end{enumerate}

% \input{preliminaries}
\section{Framework for LLMA}
% \begin{figure}
%     \begin{center}
%     \includegraphics[width=\linewidth]{figures/fr1.pdf}
%     \end{center}
%     \caption{
%         Overall framework of LLM advertisement service. \sscomment{Align $q$ and $c$.}
%         \sscomment{Remove notational things and alphabets}
%     }
%     \label{fig:teaser}
% \end{figure}

\begin{figure}
    \begin{center}
    \includegraphics[width=1\linewidth]{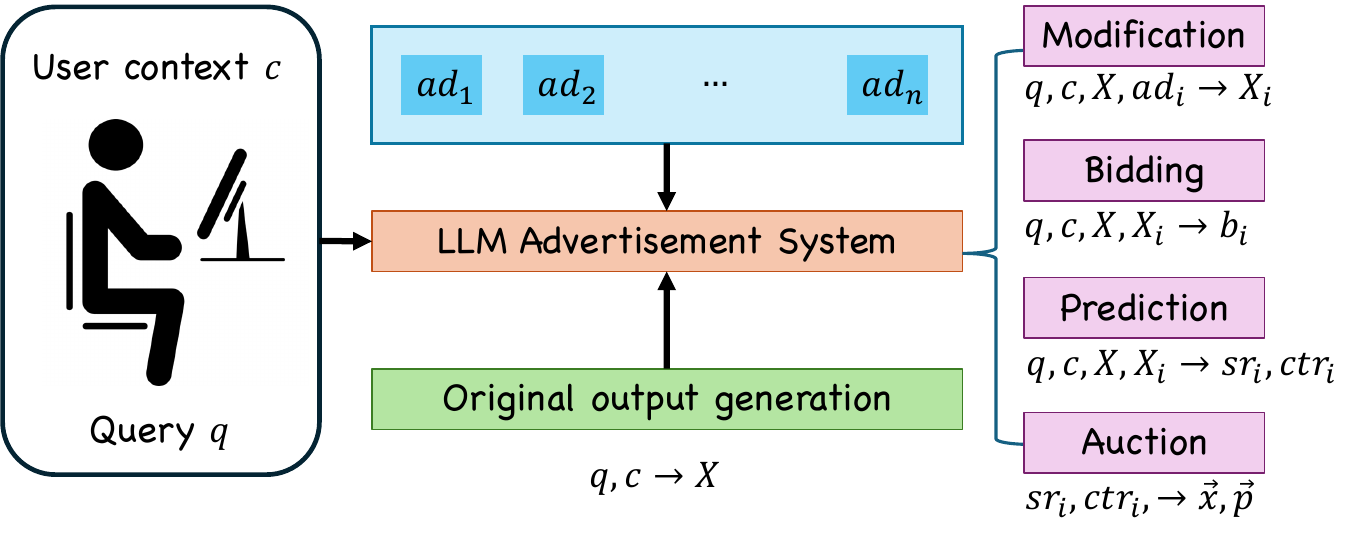}
    % {figures/teaser-fig.pdf}
    \end{center}
    \vspace{-2mm}
    \caption{
        Overall framework of LLMA.
        % \sscomment{Align $q$ and $c$.}
        % \sscomment{Remove notational things and alphabets}
    }
    \label{fig:teaser}
\end{figure}
% \vspace{-3mm}
% \SF{this is our proposed pipeline for LLMA;}
We here present a potential framework for LLMA.
We introduce several feasible types of frameworks for LLMA, each of which has its own pros and cons.
Mainly, we focus on a scenario in which a user provides a query $q$ to the LLM, and the original output by LLM is given by $X$.
Further, a~context $c$ captures a variety of features that are relevant to the advertisement recommendation, \eg history of the previous queries, user segment, region, and date.\footnote{Note here that we assume LLMA could collect such information from the user similar to the standard search engines, but our model accommodates the setting without such data. An illustrative example using such context is provided in Figure~\ref{fig:personalized-ad} in the appendix.}
Although the number of advertisers varies from time to time, when the user inputs the query $q$, we suppose that there are $n$ advertisers (bidders) indexed by $\adv_1, \ldots, \adv_n$, each of which is equipped with a single advertisement (ad) $\ad_i$ he/she wants to post.
% We denote $\adv_i$'s ad by $\ad_i$ for $i \in [n]$, and let $D = \{\ad_1,\ldots, \ad_n\}$.

Overall, we divide the LLMA as follows based on their functionalities\footnote{These functionalities may be distributed to multiple market players such as supply side platform, display side platform, or ad exchange, as does in the current online ad eco-system.}
into $4$ modules: (i) output modification, (ii) bidding, (iii) prediction, and (iv) auction.
% \begin{itemize}
% 	\item Output modification module
%     \item Bidding module
% 	\item Prediction module
% 	\item Auction module
% \end{itemize}
This overall framework is illustrated in Figure~\ref{fig:teaser}. In short, 
\begin{enumerate}
    \item The user writes query $q$ possibly with context $c$;
    \item The LLMA generates the original output $X$;
    \item Modification module creates modified output $X_i$ per ad.
    \item Bidding module generates corresponding bid $b_i$.
    \item Prediction module predicts user satisfaction rate (SR) $\sr_i$ and click-through-rate (CTR) $\ctr_i$.
    % the quality of the output $X_i$ by measuring the user satisfaction rate $\sr_i$ and click-through-rate $\ctr_i$.
    \item Auction module determines the final output and corresponding payment to charge the selected advertiser.\footnote{We first focus on presenting a single advertisement in the LLM output,
    however, generalization of the proposed framework to incorporate multiple ads at once is discussed in Section~\ref{sec:discuss-auction}.}
\end{enumerate}

% In short, the user asks a query $q$ to the system along with a context $c$, and the LLMA first generates the original output $X$.
% Then, the modification module generates specific output $X_i$ for each of the advertisers, and the bidding module generates corresponding bid $b_i$.
% At the same time, the prediction module predicts the quality of the output $X_i$ by measuring the user satisfaction rate $\sr_i$ and click-through-rate $\ctr_i$, which will be elaborated later.
% Finally, the auction module runs auction to determine the final output and corresponding payment amount to charge the advertiser.
% \sscomment{Add enumeration}
% Also, note that we do not restrict the final output to be exactly belonging to one of $X_i$, but can be totally different one based on the advertisers' bids $b_i$ and corresponding output $X_i$.
% \sscomment{fix this clearly}
For the rest of the section, we will explain each module and its responsibility/functionality in a sequential manner.

\subsection{Modification module}
The modification module takes responsibility for generating modified output based on the ads' textual information.
This module takes the pair of $(q,X,c)$ and a set of advertisements as the input, and returns $(X_i)_{i \in [n]}$ where $X_i$ denotes the modified output for $\ad_i$.

Overall, we consider two design choices:
\footnote{
For either cases, one can generate the modified output by giving an additional query to the LLM as presented in Figure~\ref{fig:intro}.
}
\begin{enumerate}
    \item In \emph{advertiser modification} model, the role of generating the modified output is delegated to each advertiser.
    \item In \emph{LLMA modification} model, LLMA directly generates the modified output.
\end{enumerate}
In Section~\ref{sec:discuss-modi}, we explore ways to reduce the computational load of generating candidate outputs.
% We further discuss how one could decrease the computational burden of generating every potential candidate output in Section~\ref{sec:discuss}.
% In either model, there would be several ways to modify the output.
% For example, one can do a prompt engineering by manually writing static sentences that are parameterized by the ads and append them to the original query.
% Alternatively, one may consider each advertisement as a document in a context of retrieval-augmented generation by~\cite{lewis2020retrieval}, and generates the final output by conditioning on the documents.

\paragraph{Comparison to SA}
Note that the standard SA does not have modification module explicitly as it is trivial to incorporate ads in each slot.
Thus, it is the unique challenge appearing in LLMA.
Further, we note that LLM is only directly used in the modification module as well as in generating the original output $X$, whereas other modules do not require LLM to operate them. Nevertheless, they entail specific technical challenges adhered to LLMA, which will be elaborated accordingly.

% \sscomment{At the end of each subsection, add a para discusses difference vs SSA.}

\subsection{Bidding module}
The bidding module generates bids based on modified outputs. It takes the query $q$, context $c$, and modified outputs $(X_i)_{i \in [n]}$ as input and outputs bids $(b_i)_{i \in [n]}$, representing each advertiser's private valuation of impressions, clicks, or conversions. We consider two design choices.

% The bidding module generates bid based on the modified outputs.
% The input to the bidding module is the query $q$, context $c$, and modified outputs $(X_i)_{i \in [n]}$, and the output will be bids $(b_i)_{i \in [n]}$ \ssedit{which represents each advertiser's private valuation on either of impression, click, or conversion.}
% We consider two design choices.
\begin{enumerate}
    \item In the dynamic bidding model, we provide the query $q$, context $c$, original output $X$,\footnote{An encrypted context $\hat{c}$ or no context could be considered for privacy.} and modified output $X_i$ to the bidder for each query, who then returns the bid.
    \item In the static bidding model, each bid is based on keywords from a pre-committed contract, without further communication with the advertiser.
\end{enumerate}
The static bidding model operates by extracting keywords from $q$, determining relevant ads, and requiring advertisers to set targeted keywords. In Section~\ref{sec:discuss-bidding}, we discuss extensions of these bidding models that allow more flexible bidding strategies. This approach can be preferred when the bid unit accurately represents the advertiser's true valuation, regardless of output quality, such as clicks or conversions.

Dynamic bidding models interest advertisers with their measures to estimate the quality of modified output. For instance, if LLMA only considers CTR and ignores user experience, advertisers may wish to adjust bids based on their output quality assessment. Despite a high likelihood of clicks or conversions, poor user experience with low-quality outputs could harm satisfaction with the advertised product.

% The static bidding model can be operated by extracting some keywords from $q$ and determines relevant ads while requiring adertisers to set targeted keyword.
% In Section~\ref{sec:discuss}, we discuss several extensions of these bidding models which allows more flexibility in customizing their bidding strategy.
% This approach could be desired when the unit of bids of interest itself well represent the advertiser's true valuation regardless of the output quality, \eg click or conversion.

% Dynamic bidding model can interest an advertiser who has its own measures to estimate the quality of the modified output.
% For example, in extreme, suppose that LLMA may only consider the CTR of the modified output and does not care about the user experience.
% The advertiser may want to customize their bid by measuring the quality of the output by itself, since he may be afraid of the system inserting its ads in a low-quality/adversarial output.
% In this case, even though the click or conversion occurs with high probability, the overall user's experience on the advertised product can be unsatisfactory due to the low quality of the LLM output.
% Thus, in comparison to the static bidding, such dynamic bidding might be desired if the advertiser wants to bid on impression while being more directly involved in the marketing.

% This approach could be desired when the unit of bids of interest itself well represent the advertiser's true valuation regardless of the output quality, \eg when the advertiser bids on click or conversion.

\paragraph{Comparison to SA}
In SA, since the context of user query is more explicit represented as keywords, advertisers can safely bid on each keyword.
In LLMA, however, it might be difficult to extract proper keywords from the query as the query itself tends to be much longer than the traditional SA due to the flexibility of LLM.
\ssedit{On the other hand, one can instruct the LLM to extract core keywords or use word embeddings to calculate the similarity between the modified output and query}.
Further, the generated output significantly affects the marketing impact of the ads in LLMA, whereas it is typically independent from other ads / contents shown within SERP in SA.
Finally, the dynamic bidding model exhibits unique challenges of dynamically adjusting the bid with respect to the output, which could further be delegated to another market player or LLM agent.

\subsection{Prediction Module}\label{sec:pred}
The prediction module computes the user's satisfaction rate (SR) and click-through rate (CTR). The SR measures user satisfaction with the output and influences the decision-making process of the language model to avoid disappointing outputs. The CTR represents the likelihood of the user clicking on the ad link in the output and is crucial for determining auction winners, as it directly impacts the LLMA's revenue. Specifically, if an advertiser's bidding method is cost-per-click (CPC), the expected revenue from the ad is calculated as CPC multiplied by the CTR.
% Prediction module is responsible for computing user's satisfaction-rate (SR) and click-through-rate (CTR).
% User's SR indicates how much the user is satisfied with the given output.
% This SR affects the final decision making process of LLMA since if there is a chance that the user is very disappointed with the modified output, then it should reconsider displaying that output as it may hurt the user experience a lot.
% On the other hand, CTR indicates the probability that the user clicks the ad link included within the modified output.
% This is also critical in determining the auction winner since this directly affects the revenue of the LLMA.
% More precisely, if an advertiser's bidding method is cost-per-click (CPC), then CTR proportionally affects the advertisement revenue as the expected revenue of transferring its ad can be computed as CPC times the CTR.
% That said, CTR may not play an important role for advertisers with bidding type cost-per-mille (CPM), since the revenue is fixed once their ads are displayed.
Overall, both of SR and CTR are functions of original output $X$, modified output $X'$, query $q$, and context $c$, which returning a real value in $[0,1]$.

\paragraph{Comparison to SA}
Different from the traditional SA whose output is static (ad image/hyperlinks), LLMA constructs textual outputs in a complicated manner. 
This makes the prediction module more difficult to learn the CTR.
Moreover, in LLMA, the prediction of user satisfaction is much more directly affected by the incorporation of the ads in the output.
This is in stark contrast with SA, whose user experience is usually affected only by the number of ad slots/positions in SERP.
Detailed methodologies for estimating/learning these functions will be discussed in Section~\ref{sec:discuss-pred}.

\subsection{Auction module}
Having computed all the required parameters, we run the auction module to determine the auction winner and the advertiser's charge. The input to the auction module is the tuples $(\bid_i, \sr_i, \ctr_i){i \in [n]}$, representing bid amount, satisfaction rate, and click-through rate. The auction module outputs an (possibly randomized) allocation $\vec{x} \in [0,1]^n$ and payments $\vec{p} \in \R_{\ge 0}^n$. Specifically, the module determines the auction format, including the allocation function, which selects the ad, and the payment function, calculating the advertiser's payment to LLMA.\footnote{This form doesn't allow for adjusting the final output to balance multiple advertisers' preferences, unlike auctions that consider bids and preferences, as in~\cite{duetting2023mechanism}. However, our framework can be extended for such cases, elaborated in Appendix~\ref{sec:discuss-comparison}.}

% Computed all the required parameters above, we finally run the auction module to determine who wins the auction and how much the advertiser will be charged.
% The input to the auction module is the tuples of $(\bid_i, \sr_i, \ctr_i)_{i \in [n]}$ which denotes bid amount, satisfaction rate, and click-through-rate, respectively.
% Output of the auction module is an allocation $a \in \{0,1\}^n$ and a payment $p \in \R_{\ge 0}$.
% Precisely, auction module determines the auction format which consists of the allocation function and the payment function, where the allocation function decides which advertisement to deliver and the payment function computes how much the advertisers pay LLMA.\footnote{In its current form, this does not allow the possibility of adjusting the final output in order to balance between multiple advertiser's preferred output.  For instance, one can think of an auction that determines the final output with respect to the bidders' bids and preferences similar to~\cite{duetting2023mechanism}.
% Our framework, however, does not preclude possibility for such extension, which will be elaborated more in Appendix~\ref{sec:discuss-comparison}.}

The main goal of LLMA is to maximize its long-term revenue by balancing short-term revenue with user retention. The objective is modeled as a function from bid amount, CTR, and SR to a nonnegative score for a modified output, i.e., selecting $i^* = \argmax_{i \in [n]} \Obj(\sr_i,\ctr_i, \bid_i)$. We do not detail the objective function choice, given the extensive literature on sponsored search auctions. After designing the score function, an auction format should be determined, with many mechanisms available, such as VCG auction or generalized second price auction as desired.

% Importantly, the main goal of LLMA is to maximize its revenue in a long term manner.
% To this end, this should balance a trade-off between short-term revenue and the user retention.
% Therefore, the objective can be modeled as a function from bid amount, CTR, and SR to a nonnegative real number that indicates a score of a modified output, \ie selecting $i^* = \argmax_{i \in [n]} \Obj(\sr_i,\ctr_i, \bid_i)$.
% We do not discuss how to choose the objective function here, given the vast literature on the sponsored search auction.
% After designing the score function, one should determine an auction format.
% Also, given the long line of literature on auction design, one can pick the most preferred mechanism such as VCG auction or generalized second price auction as desired.

\paragraph{Comparison to SA}
The main difference to the SA is, similar to what is discussed in the previous subsection, the user satisfaction is much more important measure to account for.
For example, the allocation function in SA represented by~\eqref{eq:alloc-sa} is social welfare which only accounts for platform and advertisers' utility.
In LLMA, however, one might also need to consider user's utility as a function of predicted SR, which would change the allocation function of the mechanism and the payment correspondingly.

% To briefly provide some examples, most modern online advertisement platforms use either of first-price auction (FPA) or second-price auction (SPA).
% SPA has a theoretical advantage of being incentive-compatible, thereby inducing a simple strategy of truthful bidding for the advertisers.
% Not being incentive-compatible, however, many big-tech companies run FPA for their online advertisement auctions.
% For instance, Google changes their auction format from SPA to FPA recently.
% The main reason behind such decision was that it significantly simplifies the overall advertisement process, which happened to become very complicated as lots of market players and modules came in.
% In addition, as many advertisers delegate the role of bidding process to the online ad company by using \emph{autobidder}, where the advertiser simply puts a budget threshold and then the company optimizes the bidding process for sake of the advertiser, FPA has its own advantage of simplifying this optimization procedure since the payment can be directly computed only by its bid.

\section{Market Players and Desiderata}\label{sec:require}
\ssedit{
Given the potential framework for LLMA and the presented design choices, we outline several crucial aspects for evaluating each design choice's feasibility and practicality.

In modern search or display ad auctions, numerous market players are involved, including advertisers, users, and platforms. The platform itself is often divided among several players like the demand-side platform, supply-side platform, publisher, and ad exchange. In essence, the ad platform must balance these players' utilities to create a sustainable ecosystem. For example, if the platform shows too many ads in response to a user's query, user retention will likely decrease, discouraging advertisers from using the platform, and eventually harming the ecosystem.
Similarly, for LLMA to sustain long-term revenue growth, it must balance the utilities of market players. We will outline the key aspects of the most crucial players: the user, the advertiser, and the platform.
}

\subsection{Player's incentive}
\paragraph{User experience} 
When adding advertisements to LLM output, maintaining high content quality is crucial. Users dislike excessive or irrelevant ads, which can degrade the output and reduce user satisfaction and retention. In modern online ad systems, floor prices are used to filter out irrelevant ads and preserve user experience quality. Similarly, for LLM services, we must ensure that the final output, including ads, remains a high-quality response and closely aligns with what the LLM would originally generate.
% When an advertisement is added to LLM output, it's crucial to maintain a high level of quality in the generated content.
% Users do not wish to encounter an excessive amount of advertising in the generated responses,
% especially when it doesn't align with their query or interests.
% Advertisement involvement should not significantly degrade LLM output, as doing so would lead to reduced user satisfaction, thereby decreasing the user retention.
% For instance, in modern online advertisement system, this is often handled using floor prices to filter out irrelevant ads to maintain the quality of the user experience. 
% Similarly for LLM service, we need to ensure the final modified output including advertisements can still be a high-quality response to the user, not deviating significantly from what LLM originally would generate.

\paragraph{Advertiser experience} 
Advertisers pay LLMA to include ads in outputs, expecting their products or services to be showcased compellingly. Ads should be engaging and interesting to users, efficiently driving revenue for the advertisers at smaller costs.
It is worth noting that advertisements may potentially reduce the overall number of users engaging with the system, which could have adverse effects on LLM itself.

\paragraph{Platform revenue} 
Revenue is LLMA's primary goal, so it must ensure that the additional cost of advertisements is covered by the revenue from advertisers. This is especially critical for LLMA compared to SA, due to the higher computational costs and infrastructure required to run LLMs.

\subsection{Desiderata}
\ssedit{
% To achieve the balance between the listed players' utilities for sake of a sustainable LLMA, the following criteria could be considered as desiderata from the platform's perspective.
To balance the utilities of all players and ensure a sustainable LLMA, the following criteria should be considered from the platform's perspective.
}

\paragraph{Output quality}
The LLM’s textual output should align with the user's preferences and the query. This involves (a) ensuring the ad is relevant to the user's query and (b) making sure the LLM output aligns with the user’s preferences. This includes standard LLM evaluation criteria like accuracy, relevance, coherence, and comprehensiveness.

Additionally, the output should reflect advertisers' preferences, ensuring their ads are integrated as desired and their bids accurately represent their preferences for the query and modified output.

% The textual output of the LLM should be well-aligned with the user's preference and the query.
% From the technical perspective, this can be approached in two ways: (a) allocated ad should be sufficiently relevant to the user's query and (b) given the selected ad and query, the generation of the textual output of LLM should be well-aligned with the user's preference.
% This would also include the standard evaluation criteria of LLM such as accuracy, relevance, coherence, and comprehensiveness.

% On the other hand, interestingly, this also includes the alignment of advertiser's preference into the modified output.
% In essence, it should guarantee that the advertisers' preferences on how their ads will be incorporated into the query should be well reflected, and also their bids should play as proper proxies to represent the advertisers' preferences on the query and modified output.
% should be able to reflect their preferences on the potential output by the bidding framework.

\paragraph{Allocational objective and revenue}
As a mediator, LLMA can optimize social welfare to achieve allocational efficiency, i.e., allocate ads to maximize social welfare deterministically or try to maximize its revenue.
On the other hand, probabilistic retrieval of textual information as prompts, as shown by~\cite{lewis2020retrieval}, enhances LLM performance by managing ambiguity, diversifying output, and improving robustness to noise and errors. Therefore, LLMA’s allocation objective should balance these aspects while ensuring sufficient revenue to avoid operating deficits.\footnote{This is also largely discussed in~\cite{hajiaghayi2024ad}, where we refer to Appendix~\ref{sec:discuss-comparison} for more details.}

\paragraph{Latency} 
In LLMA, users expect rapid interactions, similar to search auctions, where prompt responses are typical. Adding advertisements to LLM output introduces some latency, but this should be minimal to avoid disrupting the user experience. 
\ssedit{The latency requirement for LLMA can be less strict than for search auctions because LLMA generates output word-by-word, whereas search auctions need to retrieve all ads and results immediately upon a query.}

% In LLMA, similar to the search auction, users expect rapid interaction with LLM service, e.g., ChatGPT typically generates prompt responses, without any noticeable delay.
% Incorporating advertisement in LLM output, inevitably, will add some latency to the overall system.
% Nevertheless, this added delay should be minimal, ensuring that it does not significantly disrupt the user experience.
% \ssedit{
% On the other hand, the latency requirement could be less restrictive than the standard SA, since LLMA allows the generation of output in a word-by-word manner, whereas SA needs to ensure that all the ads and results page are retrieved simultaneously once after the user writes the query.
% }

\paragraph{Reliability and privacy}
LLMA must also address potential risks from advertisers, ensuring system reliability and alignment by considering all possible adversarial behaviors~\citep{hendrycks2020aligning}, \eg harmful contents or spam links in ads. Additionally, maintaining user privacy is crucial. All user context, information, and data must be kept secure (or encoded) to prevent privacy risks from inadvertent disclosure.
% \krcomment{I recall that in previous reviews, some reviewers said that we should explain more about this privacy? please discuss more aspects of that here.}
% \sscomment{I don't know what do discuss further.}

\section{Challenges}\label{sec:discuss}
Recall that our overall framework consists of four modules: modification, bidding, prediction, and auction module. 
For each module, we address characteristics, technical challenges, and research questions relevant to practical implementation and evaluation based on the criteria defined in Section~\ref{sec:require}. Further discussions are provided in Appendix~\ref{sec:challenge-apx}.

% For each module, we discuss several characteristics, technical challenges, and corresponding research questions in implementing the proposed architectures in practice, evaluating their overall practicality with respect to the criteria defined in Section~\ref{sec:require}.
% We provide further discussions in Appendix~\ref{sec:challenge-apx}.

\definecolor{challenge-fg}{rgb}{0.9, 0.2, 0.2}
\definecolor{challenge-bg}{rgb}{0.98, 0.5, 0.44}

% \subsection{Discussion on the requirements}
% \ssedit{On the other hand, in stark contrast to SA, most textual outputs of LLM as of now are gradually written in a word-by-word manner. Thus, different from SA, the integration of textual information on ads into the output of LLM does not necessarily need to 
% }

\subsection{Modification module}\label{sec:discuss-modi}
% \begin{tcolorbox}[colback=challenge-bg,colframe=challenge-fg]
% \begin{center}
%   Advertiser modification model that guarantees no privacy-issues, reliability, user satisfaction, and low-latency.
%   \end{center}
% \end{tcolorbox}
{\color{red}Challenges:} {\bf An advertiser modification model should ensure alignment with user preferences and satisfaction while addressing privacy, reliability, and latency issues.}
In the advertiser modification model, LLMA must provide $q$, $X$, and $C$ to each advertiser $\adv_i$, which can lead to privacy issues by disclosing user information. Addressing this involves partial or indirect information disclosure, possibly using encryption or differential privacy to protect user data while ensuring high-quality outputs.

% In the advertiser modification model, LLMA should somehow deliver $q,X,C$ to each  $\adv_i$.
% This may leak some private information of the user to the bidders, thereby suffering \emph{privacy issues}, which is critical for user experience.
% It should be addressed how LLMA can partially or indirectly disclose the information to prevent user privacy while guaranteeing high quality output from the advertiser modification module.
% Partially disclosing the information based on its level of privacy concern, using encryption, or the notion of differential privacy might be useful in this regard.

Additionally, the model faces reliability issues as advertiser-modified outputs might include illegal or spam content, which may degrade user satisfaction compared to original LLM outputs. LLMA may require an additional module to ensure robustness against such adversarial behavior, but this adds cost of computational resources and latency.

% In addition, the advertiser modification model has reliability issue since there is a possibility that the advertiser-modified output includes some illegal or spamming contents.
% The LLMA may devise an additional module to maintain its \emph{robustness} against such kind of adversarial behavior, however, this is still a cost of the LLMA to bring them in the system.
% This further possibly degrades the user satisfaction since the advertiser-generated output may not be satisfactory for the user compared to what the LLM generates.
Moreover, increased communication between LLMA and advertisers can raise latency. Therefore, developing efficient protocols is crucial for managing a functional online ad system. The advertiser modification model thus requires novel solutions to address privacy, reliability, and latency concerns.

% Furthermore, this requires additional communication between the LLMA and the advertisers, thereby increasing the overall latency.
% As such, developing an efficient protocol and algorithm to implement is necessary to operate a reasonable online ad system.
% Advertiser modification model faces many concerns as discussed, thereby requiring innovative methodologies from several perspectives to resolve all these issues.

% Overall, advertiser modification model may not be feasible in practice due to several issues discussed above.
% {\bf Research question: guaranteeing reliable, user-satisfying, low-latency output from advertiser module.}

\vspace{0.3cm}
\noindent {\color{red}Challenges:} {\bf Effectively reflecting advertiser preference on the output for LLMA modification model.}
The LLMA modification model generally faces fewer privacy, reliability, and latency issues compared to others and focuses on enhancing user experience. However, it may not fully align with advertiser preferences in the output modification process, potentially reducing advertiser satisfaction. Thus, it is important to explore methods for better incorporating advertisers' preferences into the modified output, ensuring that it meets their expectations while still improving the user experience.

% On the other hand, LLMA modification model does not suffer the same amount of privacy, reliability and latency issues mentioned above.
% Furthermore, this model would be more in favor of the user since LLMA would aim to improve the user experience.
% However, one disadvantage is that this may not fully reflect the advertiser's preference on the output modification process, thereby partly decreasing the advertiser's satisfaction.
% Thus, it should be explored how one can effectively reflect advertisers' preference on the modified output to better capture the output that advertisers would have preferred.

% This issue can be resolved by enabling more flexibility in the architecture, which will be discussed in what follows.
% {\bf Research question: well-reflecting advertiser's preference on the output.}
\vspace{0.3cm}
\noindent {\color{teal}\textbf{Prospect}:} {\bf Balancing the trade-off between LLMA and advertiser modification models.}
% Although advertiser modification model enables a~full flexibility for the advertiser to express its preference on the modification, as discussed above, it has critical issues in privacy and robustness as well as the advertiser should somehow develop its own way to modify the output.
% Meanwhile, LLMA modification model does not have these issues as LLMA directly generates the output, there is a chance that the advertisers are not at all satisfied with the modified output, thereby lowering their bids significantly.
To improve advertiser satisfaction in the LLMA modification model, one approach is to let advertisers submit indirect indicators of their preferences. Specifically, after receiving the query $q$, the original output $X$, and possibly context $c$, the advertiser provides a document $Y$ (or a list of it) reflecting its preferences. LLMA can then use $Y$ as a prompt to generate the modified output as per RAG framework, allowing more flexibility in capturing advertiser preferences. However, this method introduces additional communication costs, potentially increasing latency.

% Alternatively, advertisers could provide a list of sample documents to use as prompt engineering, which would capture their preferences without additional communication costs.

% In order to improve the advertiser's satisfaction for the LLMA modification model, one approach is to allow the advertiser to submit indirect indicators that represent its preference on what it expects the modified output to be.
% More in a detail, once the user asks a query $q$, LLMA delivers $q$, the original output $X$,
% and possibly the context $c$ to the advertiser, and the advertiser returns a string $Y$ that indicates its preference in words.
% After then, LLMA may apply $Y$ as a prompt in the LLM to generate the modified output.
% This approach allows more flexibility for the advertiser to express its preference on the output.
% Note that, however, this still requires additional communication cost compared to the LLMA modification, which possibly increases latency.
% Another way to better capture advertisers' preference on valuations without further communication is to ask advertisers to write a list of documents to give it as samples to the LLM as a prompt engineering.
% {\bf Research question: a framework that balances the trade-off given by LLMA modification, and advertiser modification models.}

\vspace{0.3cm}
\noindent {\color{red}Challenges:} {\bf Reducing the computational burden of generating every potential candidate outputs.}
\ssedit{ 
Both the advertiser and LLMA modification models require generating all possible output candidates for each potential ad. In the LLMA modification model, this involves multiple runs of the LLM’s token sequence generation for each ad in the auction. Given the high cost of LLM operations, this approach may not scale well as the number of ads on the platform increases.
% As per the description of modification module in our framework, both the advertiser and LLMA modification models require the generation every possible output candidate for each potential ad.
% In particular, in the LLMA modification model, it needs to run multiple times of LLM's token sequence generation to construct every potential candidates for the modified output with respect to the number of ads participated in the auction.
% Due to the costly operation of LLM, this approach might not be scalable as the number of ads in the platform increases.
}

\vspace{0.3cm}
\noindent {\color{teal}\textbf{Prospect}:} {\bf Ex-ante allocation without generating potential candidate outputs}
\ssedit{
One feasible approach is to implement a~prefiltering process to reduce the number of ads competing by filtering out less relevant ads. Alternatively, the auction can be run without generating every possible candidate output by using a modular component to predict the characteristics of the modified output for each candidate ad.

For instance, features can be extracted from the query and ad text, and semantic distances can be computed based on text similarity. Advertisers should be aware of this indirect measure and bid accordingly, assuming the semantic distance will represent the expected output quality.
The mechanism then uses submitted bids and semantic distances to determine the ad and generate the final output, requiring only a single generation for the allocated ad. \cite{hajiaghayi2024ad} takes this approach, elaborated further in Appendix~\ref{sec:discuss-comparison}.

% One feasible approach is to implement prefiltering process to reduce the number of ads to participate in the competition for each query by filtering out relatively irrelevant ads.
% Another approach to fundamentally deal with this issue is to run the auction without generating every possible candidate outputs by implementing a modular component to predict the characteristics of the modified output for each candidate ad.
% For example, one can consider extract several features from the query and the textual information of each ad and compute a semantic distance based on some text similarities.
% The advertisers should be aware of such use of indirect measures and write bid correspondingly.
% In essence, they assume that in expectation (over the generated output), the semantic distance would well represent the 
% Given the submitted bids and the semantic distance that captures the representative feature of the potential candidate output without its creation, the mechanism runs to determine the ad, \ie the potential candidate output.
% Then, LLM could actually generate the output given the allocated ad, thus it overall requires only a single generation of the candidate output.
% This approach is also implemented in~\cite{hajiaghayi2024ad}, where we refer to Appendix~\ref{sec:discuss-comparison} for more discussions.
}

\subsection{Bidding module}\label{sec:discuss-bidding}

% \paragraph{Comparing bidding models}
 {\color{red}Challenges:} {\bf Implementing dynamic bidding model without privacy and latency issue.}
The main advantage of the dynamic bidding model is its potential to increase advertiser satisfaction, as advertisers can adjust their bids after seeing the modified output. This flexibility can be appealing to advertisers with the technical capability to dynamically set bids.
However, the dynamic bidding model may introduce privacy issues if LLMA discloses private information to advertisers. It may also lead to additional latency since the entire set of modified outputs must be delivered to advertisers. Reliability concerns are minimal, as only the bid amount is communicated.
In contrast, the static bidding model avoids privacy, latency, and reliability issues but may reduce advertiser satisfaction because advertisers cannot adjust their bids based on the modified output.

Research should focus on the practicality of dynamic bidding, including developing protocols and algorithms that address privacy and latency concerns. 
One may also investigate how the ad market would comprise of, when there is a possibility that the advertisers hire a proxy agent to submit bids on behalf of them, and how the proxy agent (or advertisers themselves) can optimize bids in such scenario.
In addition, the emergence of autobidding system can be expected, which we detailed in Appendix~\ref{sec:challenge-apx}.

\vspace{0.3cm}
\noindent {\color{teal}\textbf{Prospect}:} {\bf Balancing the trade-off between static/dynamic bidding models.}
% \paragraph{Beyond static and dynamic bidding}
% \sscomment{change title}
In the static bidding model, improving advertiser satisfaction can be achieved by defining a more flexible static function as the contract between the advertiser and LLMA. Specifically, LLMA could propose a contract where bids are determined by an indirect measure of the modified output.
For example, LLMA and the advertiser might agree on a contract where the bid is inversely proportional to the similarity distance $d$ between the original output $X$ and the modified output $X_i$. If $X_i$ significantly deviates from $X$, the similarity distance will be large, leading to a lower bid from the advertiser due to concerns about user experience.

Advertisers will need to understand how LLMA estimates and defines this distance measure. LLMA could also develop a more refined method for assessing user interest, attention, and relevance for the modified output from the advertiser's perspective, allowing advertisers to choose contracts based on their preferences.
% Exploring these aspects is crucial for to operate
% effectively implementing LLMA in practice.

% In the static bidding model, a natural approach to improve the advertiser's satisfaction is to enable a more generic static function as the contract between the advertiser and LLM provider.
% Specifically, LLMA can propose the bidding contract as a function that maps an indirect measure of the modified output to a nonnegative real number representing the bid, \ie the advertiser's valuation.
% For instance, LLMA and an advertiser agrees on a inverse proportional contract based on the similarity distance $d$ between the original output $X$ and the modified output $X_i$.
% For example, if $X_i$ is examined to be very different form $X$, then $d(X,X_i)$ will be large, and then the advertiser's valuation, \ie bid, on $X_i$ is small since it is believed that the incorporation of ad too much changed the content of the output, thereby hurting user experience and capturing less user attention.
% Note that in this case, the advertiser may need to partly understand how LLMA estimates the distance measure and how it's defined.
% More generally, LLMA might construct a more refined way to measure the user's interest, attention, relevance for the modified output from the perspective of $\adv_i$, and let the advertisers commit to either of the contract based on its taste.
% These are crucial aspects to explore to run LLMA in practice.

\subsection{Prediction module}\label{sec:discuss-pred}
{\color{red}Challenge:} {\bf Efficient and precise implementation of prediction module.}
% To recap, the input to both the prediction modules is the original output $X$, query $q$, context $c$, and the modified outputs $(X_i)_{i \in [n]}$.
Estimating CTR in LLMA can follow principles similar to those in modern online advertising. We can train a prediction system to estimate $\ctr_i \in [0,1]$ based on input $X_i$, $q$, and $c$, using user feedback data.
For example, factorization machines and online algorithms can be used after feature extraction from user data, context, and queries, as described in \cite{mcmahan2013ad}. Given the sparse frequency of each $X_i$ in the family of possible outputs, features from $q$, $X_i$, and $c$ should be extracted to map to CTR values. User actions like regenerating responses, clicking ads, or exiting the LLM can be used to refine the prediction module.

% The process of estimating CTR will be analogous to that in the modern online advertisement system in philosophy.
% We can train a prediction system to predict $\ctr_i \in [0,1]$ given the input $X_i, q, c$ for each $i \in [n]$ based on the feedback data from the user.\footnote{For instance, as is typical in the modern online advertisement system, one may use factorization machine based online algorithm after extracting some features from user, context, and the queries. More detailed discussion can be found in \cite{mcmahan2013ad}.}
% As discussed in Section~\ref{sec:pred}, since $X_i$ will be very sparsely appear if the entire document is considered as an embedding, we may extract some useful features from each $q, X_i$ and $c$, and learn the mapping from features to the true value.
% Whether the user regenerates the response, clicks the ad, or exits the LLM, can be used as feedback data to improve the prediction module.

% Feedback data will be the direct action of the user indicating whether they click the ad or not.\sscomment{awkward sentence}
% This prediction system can be trained in an online manner to update the estimate in (almost) real time, which will increase the accuracy of the prediction.
% \sscomment{move the footnote to the main body?}
% {\bf Research question: }
\vspace{0.3cm}
\noindent {\color{red}Challenge:} {\bf Relevance/similarity distance measure to estimate user satisfaction.}
To estimate SR, one approach is to assume that the original output $X$ is optimal. In this case, the distance between $X$ and the modified output $X_i$ can serve as a measure of SR, since the closer $X_i$ is to $X$, the higher the expected user satisfaction.
For example, one might define the output distance as $d(X, X_i) := ||\text{Pr}(X | q) - \text{Pr}(X_i | q)||$ using a suitable norm. Here, $\text{Pr}(X | q)$ represents the marginal probability of $X$ given $q$, which can be computed using standard methods from the literature~\citep{vaswani2017attention}.

More general functions, such as semantic similarity between documents or phrases~\citep{mikolov2013efficient, cer2018universal, conneau2019unsupervised}, can be used. 
One can further implement a calibration layer to maintain well-calibrated with higher accuracy~\citep{mcmahan2013ad}.
The key research question is to identify effective measures for predicting user satisfaction when ads are incorporated into the output.

\vspace{0.3cm}
\noindent {\color{teal}\textbf{Prospect}:} {\bf Incorporating distance measure and online learning.}
As discussed, one may consider combining similarity measure and online learning from user feedback for prediction.
To online learn SR (or CTR) estimates from user feedback, a useful indicator of whether the user is satisfied with the output is whether the user \emph{re-generates} the output.
One may aim to learn a function which outputs the $\sr$, given the query $q$, modified output $X'$, and context $c$.
This approach does not assume that the original output $X$ is indeed optimal, thereby allowing the possibility that the user may be satisfied with a modified output $X_i$ even though its distance from $X$ is measured is large.
This comes at the cost of additional modular component for learning process.
% Note that one should implement an independent training system to operate this learning process, leading to additional burden for LLMA.
One could further consider an online learning model where the similarity distance is also integrated as one of the features for prediction.
If the similarity distance has some positive correlations with the true user satisfaction rate, this would increase the accuracy of the prediction.
Essentially, an effective way to capture the both advantages of online learning and distance measure should be studied thoroughly.

\subsection{Auction module}\label{sec:discuss-auction}
% \paragraph{Adjusted}
{\color{red}Challenge:} {\bf Incorporating multiple ads in a single output.}
% \paragraph{Multiple ads at once}
Recall that our framework is presented for single-ad allocation setting.
% In our auction module, the modified output is fully determined by a single ad.
\ssedit{
One approach to generalize our framework is to repeatedly run the overall procedure and allocate a single at once.
That is, one can determine an abstract unit that partitions the output, \eg paragraph, and run our framework for each unit.
}
Another direct approach to extend our framework is, to let the final output to be displayed the one that does not necessarily belong to $\{X_i\}_{i \in [n]}$, but rather a new output $X'$ that interpolates $\{X_i\}_{i \in [n]}$.
This resembles the approach of aggregating the preference by~\cite{duetting2023mechanism,soumalias2024truthful}.
More discussions can be found in Appendix~\ref{sec:discuss-comparison}.
% \footnote{We discuss further comparison to the token auction model by~\cite{duetting2023mechanism} in Appendix~\ref{sec:discuss-comparison}.}
% Let us denote the new output $X'$ by \emph{balanced output}, \eg see Table 2 in~\cite{duetting2023mechanism}.
By doing so, it might be possible to deliver multiple advertisements in a fair manner, thereby allowing LLMA to bring more revenue by charging multiple advertisers at once.
\footnote{Another way to display multiple ads at once will is discussed in Appendix~\ref{sec:challenge-apx} under our original framework.} 

One subtle issue is that, since each advertiser bids $b_i$ on delivering $X_i$, they may not want to write the same bid for the balanced output $X'$, thus it may degrade the advertiser's experience.
In the static bidding model, as discussed in Section~\ref{sec:discuss-bidding}, this might be handled by committing to a contract based on measures that represents the advertiser's preferences more in a refined manner.
In the dynamic bidding model, one approach would be to append an additional step of asking for bids for the final output again to the advertisers.
% In summary, incorporating multiple ads in a single response is technically challenging, which would require both theoretical and practical studies to validate reasonable approaches.

\subsection{Discussion on Recent Approaches}
\ssedit{
Several recent theoretical approaches have proposed game-theoretic models for ad auctions in LLMs. 
We here discuss how these can be seen as implementations of our framework.\footnote{Further details are deferred to Appendix~\ref{sec:discuss-comparison}.}

First, \citet{duetting2023mechanism} propose a model where bidders submit bids and distributions over tokens. This aligns with a tokenized version of our model with (i) LLMA modification by aggregating token distributions, (ii) dynamic bidding with adjustments for each query, (iii) no prediction module as it focuses on advertiser perspectives, and (iv) running a token auction.
This approach might suffer from latency and reliability issue, but could possibly maintain high advertiser experience.

\cite{hajiaghayi2024ad} integrate the RAG framework for segment-based ad auctions, where ads are probabilistically retrieved for segments like paragraphs using bids and a notion of relevance. This corresponds to (i) LLMA modification without pre-generating outputs, (ii) static bidding based on clicks, (iii) indirect CTR prediction by measuring ad relevance, and (iv) running a segment auction.
Their approach has less privacy, latency and reliability, but advertiser's experience could be largely dependent on how modular components to compute relevance is designed.

\citet{dubey2024auctions} introduce a framework that generalizes position auctions by determining payment and a notion of prominence as allocation function. This aligns with (i) LLMA modification, (ii) static bidding, (iii) no prediction required, as prominence serves as a CTR proxy, and (iv) running a prominence auction.
Similar to\cite{hajiaghayi2024ad}, advertiser's experience highly depends on how LLM well constructs the output with desired prominence.

\citet{soumalias2024truthful} propose an auction that truthfully aggregates advertiser preferences using RLHF framework. This can be viewed as (i) LLMA modification, (ii) static bidding, (iii) no prediction needed, focusing only on advertiser perspectives, and (iv) running an RLHF-based auction.
This requires relatively larger latency issue since every potential candidate output needs to be created and advertiser's reward function should be communicated accordingly, however, it could possibly reflect advertiser's preference better.

}

\section{Conclusion}
This paper addresses the opportunities and challenges for leveraging an online advertisement system with LLM.
We first discuss essential requirements which a reasonable LLM advertisement system should satisfy.
To fulfill such desiderata, we introduce several frameworks  and substantiate their validity.
We compare proposed frameworks with their advantages and limitations.
Then we discuss further technical and practical challenges in designing efficient online advertising with LLM.
Finally, as a usecase of independent interest, we present the potential of advanced dynamic creative optimization by using LLM.

% \section{Impact Statement}
% Our paper aims to explore the potential of online advertiseemnt system with LLM and suggests several of its plausible directions in practice. The potential broader impact of our research encompasses several dimensions. Firstly, it advances the practical field of machine learning by initially suggesting several plausible approaches to run online advertisement system for LLM-based search engine. Secondly, our work has implications for the online advertising industry itself, potentially leading to more personalized and targeted advertising campaigns that better align with user interests and preferences.

% Ethically, the deployment of LLM-based advertisement systems prompts considerations regarding user privacy and content reliability. We acknowledge the importance of addressing these ethical concerns and emphasize the need for responsible implementation practices and mechanisms.

% Furthermore, the societal consequences of our work include shaping the digital landscape and influencing user experiences on online platforms. As online advertising plays a significant role in shaping consumer behavior and influencing purchasing decisions, our work may impact the dynamics of online commerce, search engine and corresponding consumer engagement.

% \section{Impact Statement}
% \sscomment{To be done}
% \input{ack}
\bibliographystyle{plainnat}
\bibliography{ref}
% \newpage
% \input{aaai2025/reproduce}
% \newpage
\newpage
\appendix
\onecolumn
% \section{Further Related Works}\label{apd:further-rel}

\section{Further Related Works}\label{app:fur-rel}
\paragraph{Chatbot advertisement}
% \sscomment{Skip this or move to the appendix}
With emergence of chat-bot in various websites,  several companies started to run a chat-bot advertisement, which is often referred to as sponsored message.
In the chat-bot advertisement, the companies send some private messages on the chat-bot conversation to advertise some products.
This is very different from operating an~online advertisement system over LLM, since chat-bot advertisement simply sends some messages, but our model captures the scenario in which the user asks a query and then LLM provides a output including some advertisements.
Notably, Microsoft Bing started to operate an online advertisement with Bing chat-bot (\eg see articles ~\cite{Adweek, Techcrunch, Microsoft}).
Indeed, once the user inputs a query in the chat-bot, the output often includes advertisements with explicit flags that indicates this is an advertisement.

\paragraph{Dynamic creative optimization}
Dynamic Creative Optimization (DCO) is an online advertising strategy that involves dynamically tailoring the content of an ad based on user behavior, demographics, and user context~\cite{baardman2021dynamic}. 
Instead of displaying a static, one-size-fits-all ad, DCO allows advertisers to create and serve personalized and relevant ad content to different audience segments.
By doing so the platform can improve the likelihood of user engagement, click throughs, and conversions.
DCO typically involves the use of technology and algorithms to automate the process of selecting and assembling ad elements, such as images, copy, and calls-to-action, in real-time. This customization is done based on user data, collected either in real-time or from previous interactions. 
Unlike DCO has been typically used only in display advertising not search advertising~\cite{choi2020online} since the tailored ad creatives, \eg personalized hyperlinks to websites, would not have much impact for structured output of standard search engines.
Due to the unstructured nature of LLM advertisement, the use of techniques similar to dynamic creative optimization would be highly likely to be present in future.
Our paper suggests the use of LLM, more generally generative AI, to personalize the advertisement itself  to improve the relevance and quality of the advertisement.

\paragraph{LLM-assisted Ad Creation}
Beyond using LLM for DCO, one may use it to implement a system that can help the advertisers generating the ads itself.
Indeed, there already exists a tons of online platforms such as Japser AI for copywriting, Lexica Art for blog thumbnails, and Attentive for mobile messaging, to help marketers construct their static ads.
Further, the multi-modal capability of the very recent ChatGPT-4o by OpenAI opens the possibility of supporting advertisers creating ad assets with text-based interactions.
In addition, Meta recently introduced an AI sandbox in their~\cite{MetaAdvantage} that can help advertisers from several perspectives, including creating a variant of the given ad, constructing an ad based on text prompts or quickly changing the ad format with respect to specified ad channels~\citep{MetaSandbox}.

\begin{figure*}
    \begin{center}
    \includegraphics[width=\linewidth]{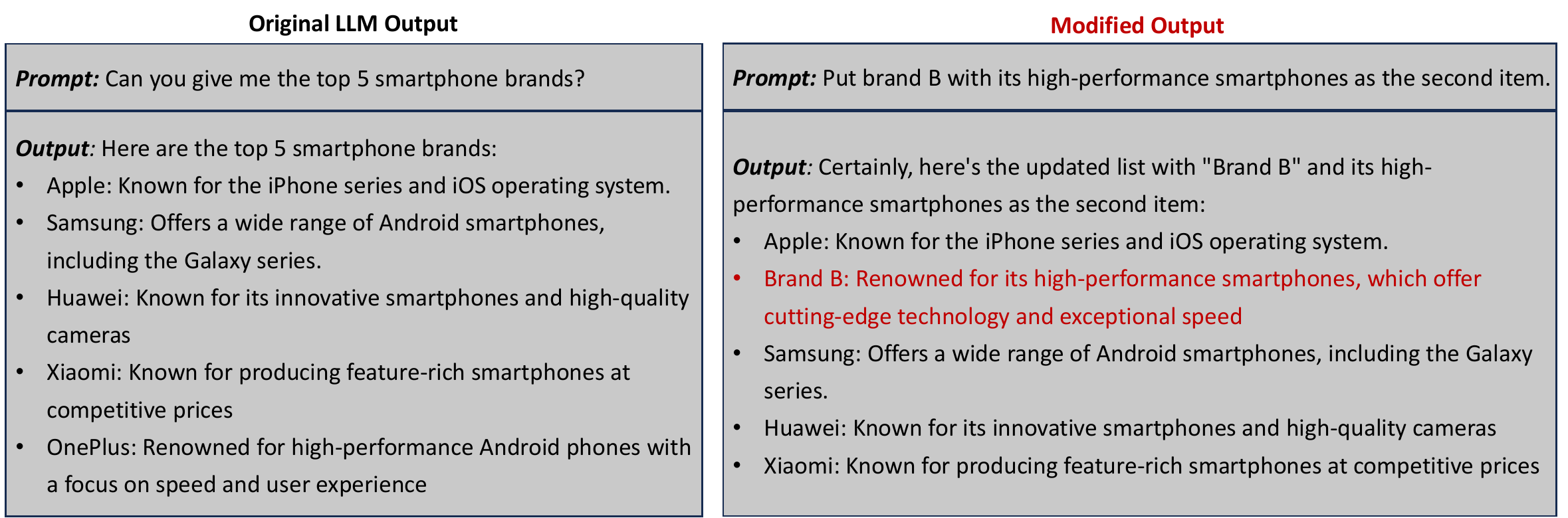}
    \end{center}
    \caption{An example of providing \textbf{structured} advertisement in the LLM. Left column refers to the case where we ask above question from ChatGPT 3.5,
    and to \textbf{incorporate the ads} we use queries specified in the right column.}
        
        % In the \textbf{right panel}, the user is interested in machine learning conferences in Florida.
        % Chat bot is advertising ``SunShine Travel" which is a fictional travel agency by putting its logo on a general image from Florida (left)
        % Incorporating personalized advertisement, model can generate more specific advertisements by offering flight tickets
        % from the place that the users interacts with the system and showing a more related image of Florida.
        % For example, when a young unmarried person from DC interacts with the system,
        % output image could advertise tickets from DC to Florida and could be modified to more appealing for the user (middle)
        % or when a married person with their family living in California enters the query, the image could promote flight tickets from there to Florida
        % while showing an image of amusement parks in destination.
    
    \label{fig:intro2}
\end{figure*}

\begin{figure*}
    \begin{center}
    \includegraphics[width=\linewidth]{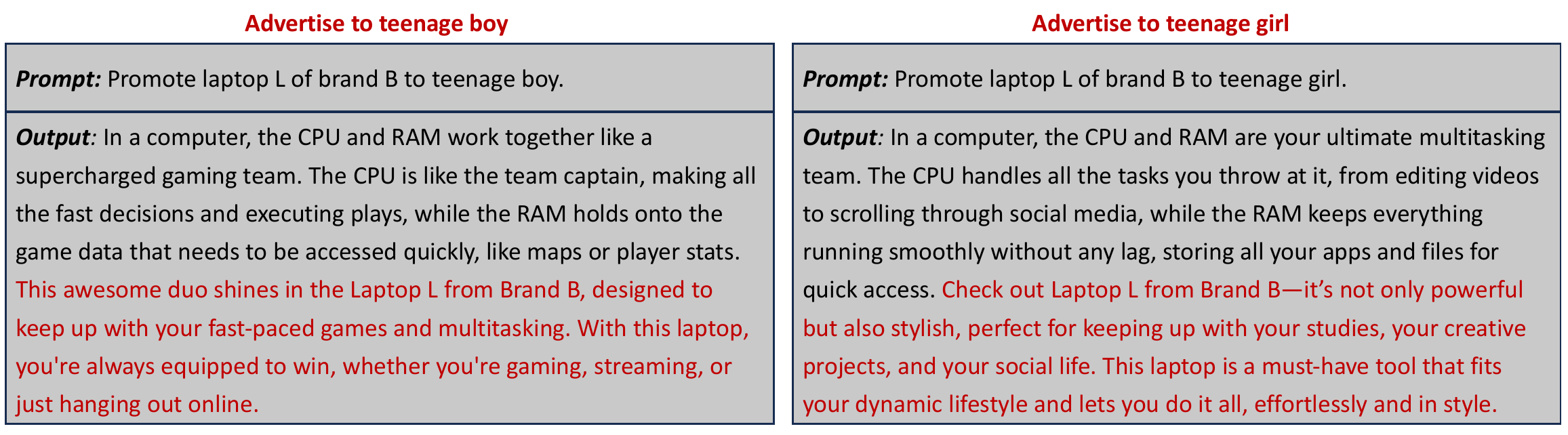}
    \end{center}
    \vspace{-2mm}
    \caption{An example of ChatGPT 4 providing different output tailored to the user context. We keep the context provided in Figure~\ref{fig:intro}, but asks again to advertise again to the specified user segment.}
    \label{fig:personalized-ad}
\end{figure*}

\section{Recent Works on Mechanism Design for LLMs}\label{sec:discuss-comparison}

\paragraph{\cite{duetting2023mechanism}}
% \krcomment{I believe this paragraph could be pushed to Appendix. Only a short gist of the idea can be discussed here. We are writing a position paper, don't need to discuss these details.}
The approach we discussed at the end of Section~\ref{sec:discuss-auction} (Auction module part) is similar to the \emph{token auction} model proposed in~\cite{duetting2023mechanism}, under which the advertisers equipped with their own LLM directly provide a desired distribution over the subsequent token and corresponding bid, and then the mechanism finally decides the final distribution by running a token auction.
They provide a methodology to aggregate distribution and commits to the final one, while constructing a reasonable payment scheme to respect the reasonable properties of auction in the mechanism design perspective.
Importantly, however, their model does require the advertisers to bid on their preference for each token, thereby requiring more communication cost to construct the entire output.
Our advertiser modification and dynamic bidding model can be thought as a version of their model, but the advertiser only bids on the final deterministic output.
In this context, one may consider generalizing this model by letting the advertisers to bid on a distribution over the final output, thereby allowing the advertisers to represent their preferences more in a refined way.
Note, however, that since the entire space of possible output may be extremely large, let the advertisers bid a distribution over such space might be practically/computationally infeasible.
Alternatively, one can consider a version in which each advertiser provides a potential set of outputs, (partial) ordering over it with corresponding bid, and then the auction module determines the final output based on designed mechanism.
\ssedit{
Overall, their approach can be viewed as (i) LLMA modification that aggregates the preference reported by advertisers and (ii) dynamic bidding model where the advertisers bid on their desired preference.
}

\ssedit{
\paragraph{\cite{hajiaghayi2024ad}}
Recent work by~\cite{hajiaghayi2024ad} considers the integration of RAG framework into the ad auctions for LLMs. In essence, they consider a single-dimensional mechanism design problem where each advertiser has a private value on each click, and a notion of \emph{relevance} between the query and the ad's textual information the plays as a proxy for the click-through-rate of an allocated ad in the final output.
Formally, they assume that the module to calculate the relevance is calibrated to be proportional to the expected click-through-rate over the possible outputs given the context of ad and query.
Mainly their mechanism determines the allocation and payment function given the bids and relevance of advertisers.
Precisely, their mechanism first computes a perturbed score inspired by discrete choice methods~\cite{train2009discrete},
randomly retrieves an (or a set of) ad with respect to the probabilities given by RAG framework, and charges the ad based on the perturb scores.
Interestingly, they take an approach of reverse-engineering the objective function that the allocation function given by RAG framework tries optimize.
They show that the linearly probabilistic allocation from RAG in hindsight maximizes logarithmic social welfare.
From our LLMA framework's perspective, their approach can be viewed as the combination of (i) LLMA modification but without generating all the potential candidate outputs as discussed in Section~\ref{sec:discuss} (Modification module part), (ii) static bidding, (iii) CTR prediction is indirectly done by measuring relevance and (iv) running an auction with perturbed scores.

\paragraph{\cite{dubey2024auctions}}
\cite{dubey2024auctions} considers an ad auction framework for LLMs as a generalization of position auction in the standard SA.
Similar to the relevance in~\cite{hajiaghayi2024ad}, they introduce a notion of prominence that captures how much each ad takes portion in the output. 
In their approach, each advertiser writes a single-dimensional bid, given a collection of features for each ad, their mechanism determines the prominence of each ad within the output.
Alternatively speaking, they calculate a desired CTR for each ad as a mechanism's output.
Then, their LLM module retrieves the calculated prominence for each ad as an input, and it generates the entire output that is consistent with the calculated prominence.
If one deems the positional effect in the standard position auction used in SA as the prominence, their approach can be viewed as a generalization of the position auction.
They formalize such correspondence, and characterize several theoretical properties of their prominence auction.
One subtlety in their framework is that LLM might not reliably produce an output that is exactly consistent with the desired output.
In our framework's language, their approach can be viewed as (i) LLMA modification, (ii) static bidding, (iii) CTR is not required for the payment but instead prominence plays this role, and (iv) auction module calculates the prominence required for LLM to generate the output.
% mechanism retrieves prominence (\ie desired CTR) of each input and generates the 

% Note that their approach does not consider a user's query as an input, since their primary objective is to design a mechanism to summarize several textual information in an auction format.
% However, the integration of the query into their framework naturally can be accommodated 

\paragraph{\cite{soumalias2024truthful}}
Finally, \cite{soumalias2024truthful} proposes an auction framework to truthfully aggregate the advertisers' preferences.
Each advertiser's preference can be represented via its own LLM or a reward function.
Given multiple advertisers' reward functions, the mechanism's objective function is to maximize the summation of the advertisers' rewards minus a normalization term to not too much deviate from the reference output with no advertisement.
They observe an intricate connection of such objective function to the reinforcement learning from human feedback (RLHF) approach by~\cite{ziegler2019fine}, and use these devices to construct a dominant-strategy incentive-compatible mechanism.
From our perspective, their approach can be viewed as (i) LLMA modification, (ii) static bidding, (iii) prediction module is not required as the advertiser's preference is completely determined by the output, and (iv) an auction with allocation function based on the RLHF with aggregated reward functions.
}

\section{Further challenges and practical concerns}\label{sec:challenge-apx}
\paragraph{Autobidder}
\ssedit{
In the modern search ad or display ad auction in ad platforms, significant amount of advertisers use an autobidding system provided by the ad platforms to delegate the role of dynamically optimizing the bids with respect to the user.
The autobidding system, usually referred to as autobidder, is even more crucial for advertisers who do not have sufficient budget to run a market research to determine the appropriate amount of bids.
In autobidder, each advertiser only is required to determine a high level objective it wants to achieve.
For instance, they may specify that they want to maximize the number of clicks subject to the budget constraint of $\$100$ per day, or that they want to spend at most two bucks per each click on average.
The former is typically referred to as autobidder with the budget constraint, and the latter as with the return-on-investment (ROI) constraint.
Due to its centralized nature, autobidding intrinsically enables the platform to dominate the decentralized bidding scenario in social welfare as well as the platform's revenue.
Correspondingly, a long line of works~\citep{lucier2024autobidders,feng2024strategic} aim to explore an efficient autobidding algorithm, analyze the dynamics of the platform with autobidding and manual bidding advertisers, identify the resulting outcome of each auction format under autobidders.

In a similar vein, we expect that autobidder could also play an essential role to optimize the markets in several directions for LLMA as well.
The platform may allow the advertisers to specify budget or ROI constraint as in the standard search or display ad auction, and the set of feasible objective functions as impression or click.
Due to more complicated nature of ad appearance in the output, it might allow alternative objective functions compared to the standard autobidding system such as number of sentences or paragraphs dedicated to the ads, etc.
The platform can also have broader optimization space over bids sequences since it can dynamically adjust how much portion it would spend on the output for each ad and corresponding ad.
As the autobidder system inevitably emerges as a solution in the modern search ad auction, we expect that a similar device, but perhaps with a complicated design, would necessarily emerge in ad auctions for LLMs.
}

\paragraph{Agentic Workflow}
\ssedit{
One notable approach to implement each module of LLMA is to construct a dedicated LLM agent to help serve the module in a similar vein with Agentic Workflow~\citep{madaan2024self}.
For instance, in the LLMA modification model, LLMA might implement an LLM agent specialized to each segment of the ads to generate the ad-incorporated output.
For the prediction module, it could exploit an LLM agent that can perform a feature extraction task to deliver useful features from the user's query and possibly context to the prediction module.
This agent could further collaborate with another LLM agent that can compute the predicted CTR and SR or any other dedicated machine learning module.
Even for the bidding module, from the autobidding perspective presented above, LLMA could utilize an LLM agent that determines each ad's bid on behalf of the advertiser. As the use of such dedicated LLM agent might be costly, there could be several options for contract between LLMA and advertisers, \eg whether the advertiser wants to use an entirely dedicated agent specialized at higher cost or a cheaper general-purpose agent.
These LLM agents do not necessarily solve the entire task at hand, but could greatly improve the efficiency of each module once equipped with a proper prompt engineering.
}

\paragraph{Parallelizing the process and adapting multiple types of advertisers}
Recall that our overall architecture consists of (i) modification, (ii) bidding, (iii) prediction, and (iv) auction modules.
Although each of the process should be serialized in order to handle a single advertiser's process, it is important to observe that the process (i)-(iii) is totally independent between the advertisers.
Hence, one efficient to way to minimize the latency is to process steps (i) to (iii) in an asynchronous manner between the advertisers.
To fit the desired latency requirement, one may consider excluding some advertisers such that the steps (i) to (iii) are not completed within some maximal delay. It's common to have timeouts in display ads, because bids can come from real-time bidders outside the system, but not as much for search ads as far as I'm aware.
After collecting all the feasible parameters, LLMA can pass the required parameters to step (iv) and run the auction.

One additional advantage of this procedure is that, one can easily adapt to multiple types of advertisers.
In practice, several advertisers may want to dynamically bid or modify the output by themselves, while others with small businesses may not want to take too much of responsibilities.
For instance, consider the following scenario.
\begin{enumerate}
    \item $\adv_1$ wants to modify the output by itself, and dynamically bids on it.
    \item $\adv_2$ does not want to modify the output by itself, but wants to dynamically bid.
    \item $\adv_3$ does not want to modify the output by itself, and wants to bid in a static manner.
\end{enumerate}
If one conceptually separates the steps (i), (ii) and (iii) between the advertisers, we can adapt all of these advertisers by running a separate procedure for each advertiser, and then finally runs the step (iv).

\paragraph{Displaying multiple ads simultaneously}
We may want to make it clear in the intro that the paper focuses on displaying a single ad, and that multiple ads bring their own challenges discussed briefly at the end.
One common feature in the modern sponsored search auction is that, it usually display multiple ads upon a single query, which is mostly to maximize the search engine's revenue.
Given the architecture we proposed, it is not straightforward to allow this feature, so we here discuss several approaches to deal with it.
% We will start from a simplistic scenario, and then gradually complicates it.

First, consider a simple case in which all advertisers use LLMA modification and static bidding model.
In this case, one can consider every combination of advertisements as a subset of $[n]$, and then generate modified output with corresponding bid.
Further, if there is an advertiser with dynamic bidding, LLMA can simply delegate the bid generating process to that advertiser for every combination that includes this advertiser's ad.

Obviously, these approaches are not computationally feasible if one consider every possible combinations of the subset.
% This is even more problematic if there exists an advertiser with dynamic bidding since the computational burden to communicate the collection of modified outputs can be extremely heavy.
Therefore, the technical challenge lies in how one can optimally decide a subcollection of the entire power set to balance the trade-off between computational burden with corresponding latency and revenue of the auction.

% \sscomment{
% \paragraph{Personalized ad creation}
% In case the advertisers use LLMA modification model, one further improvement that LLMA can incorporate is to personalize the advertisement messages or figures based on what the advertise provides as ad creatives.
% The gains from this procedure might be more significant if the modified output contains advertisement as a figure inside the output.
% For example, suppose that an advertiser who wants to advertise its brand new desktop products with several models.
% Consider a user who asks LLM about how a desktop computer operates overall.
% LLM may provide an illustrative figure to describe how a computer operates in general.
% Depending on the user's context, \eg gender/location/the current device used for the query, it is possible to personalize the advertisement by incorporating the suitable model that the user may prefer the most.
% This procedure can not only increase the revenue of the advertiser and LLMA, but also improve the user experience.
% }

\section{Dynamic Creative Optimization via LLM}\label{sec:personalized-ad}
\begin{figure*}
    \begin{center}
    \includegraphics[width=\linewidth]{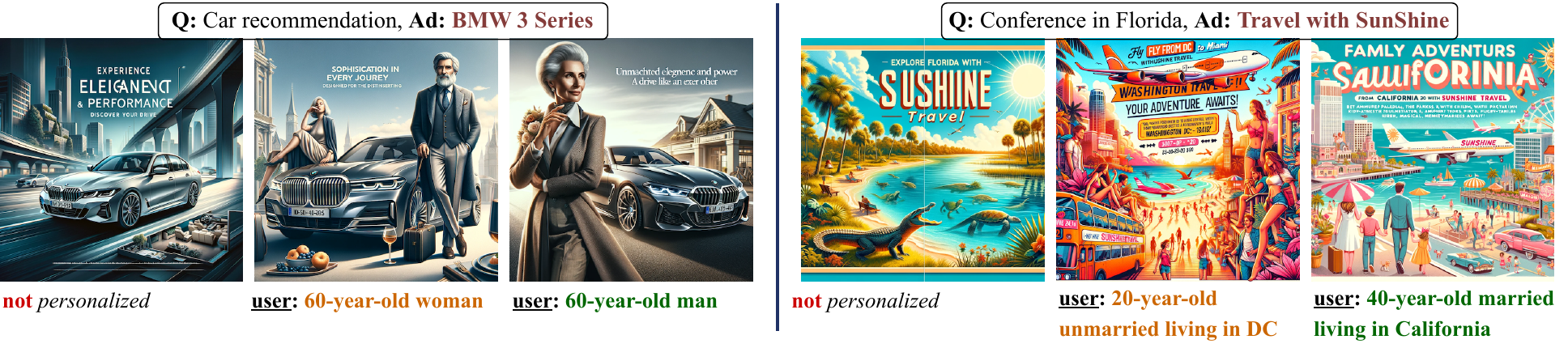}
    \end{center}
    \vspace{-3mm}
    \caption{
        Integrating \textbf{responsive advertisement} in the images generated by ChatGPT 4.
        In the \textbf{left panel}, a user requests car recommendations within the price range of \$40,000 to \$60,000, 
        and the chat bot, when capable of advertising, generates a ``BMW 3 Series" image (left).
        Personalized ads modify this image for a 60-year-old woman (middle) or man (right) based on user context like age and gender.
        Note that it includes an image of woman for man in the output, and vice versa.
        In the \textbf{right panel}, a user seeks machine learning conferences in Florida. The chat bot advertises ``SunShine Travel" with its logo on a generic Florida image (left).
        With personalized ads, the model tailors promotions, offering flight tickets from the user's location and showing more relevant Florida images.
        For example, for a young unmarried person from DC, the output advertises DC to Florida flights (middle) and features an image of beach there,
        while for a married person in California, it promotes California to Florida flights with images of amusement parks.
        This \textbf{enhances user experience} with targeted and appealing content.
        }
    \label{fig:personalized-ad}
\end{figure*}

Now, we discuss about the possibility of advanced dynamic creative optimization by using LLM.
\emph{Dynamic creative optimization} (DCO)
\footnote{This is often called as dynamic product ads (DPA), but DPA typically refers to the process of selecting which item to advertise from a given catalog of an advertiser, whereas DCO denotes the process of modifying the ad itself.},
in the traditional search and display ad markets, refers to the technique of dynamically adjusting the combination of ad assets and constructing an advertisement that best fits the customer's preference.
It can markedly enhance the quality and relevance of the ad to the user by tailoring the contents of ad in a personalized manner.
Many online advertisement platforms adopt DCO-based ads, which is often called as \emph{responsive ad}~\citep{GoogleAdSearch,GoogleAdDisplay}.
\footnote{As traditional DCO is restricted to dynamically construct the complete image file of ad, we here restrict our attention to the image-based ads as well. However, it is apparent that LLM based DCO would work beyond image-based ads.} 
% \krcomment{Fantastic footnote!}}

In the canonical DCO framework, a static ad consists of a single image file that can only be displayed in a specific set of ad channels, whereas dynamically optimized ad consists of a single ad template and multiple options for each asset that fills in the ad template.
This is typically done by the advertiser registering multiple options for each asset given an ad template, and then the system picks the optimal combination.
% \footnote{Note that this is different from picking and delivering the most relevant ad given a pool of ads, but about tailoring the ad itself to satisfy the user.}
This procedure aims at increasing the CTR and SR, which would eventually improve the welfare efficiency of the online ad ecosystem~\citep{mcmahan2013ad} as well as the revenue of the ad platform.

% It's important to recognize that tailoring advertisements to individual preferences can enhance their quality, making them more appealing to users.
% Consequently, the user's satisfaction rate tend to rise, accompanied by an increase in the click-through rate when exposed to personalized advertisements.
% This positive correlation leads to higher revenue for both LLM provider and advertisers, aligning with the goals of the effective advertisement.
% This brings more revenue for LLM provider as well as advertisers and improves user experience, thereby increasing the welfare efficiency of the online ad ecosystem~\cite{mcmahan2013ad}.

Unlike from the traditional DCO techniques, the arise of LLM, especially those support the output with image such as ChatGPT 4, enables more involved process of dynamic creation of ads, as well as replaces the role of traditional DCO in online ad platforms.
For example, given a user query and context, the platform may send a query to LLM to customize the image of ad to attract the user by capturing user preference.
We provide representative examples in Figure~\ref{fig:personalized-ad} where two scenarios of personalizing ads is depicted.
Note that the use of LLM-based DCO is not restricted to LLMA, but can be broadly applied to SA and display ad.

\vspace{0.3cm}
\noindent {\color{red}Challenge:} {\bf Refining each module.}
% \paragraph{Incorporating responsive ad}
To integrate responsive ads, a more advanced modification module is imperative.
This module should not only enable the alteration of the original response to include ads but also incorporate them in a manner that takes into account user preferences.
This entails utilizing user contexts, \eg gender/location/the current device used for the query, in the ad generation process.
To be more specific, the modification module should incorporate user context in the process of modifying the original response.
This makes ads more appealing to the user by considering their preferences.
Leveraging language models, we can prompt them to factor in various elements related to user context during the generation of modified outputs.
As illustrated in Figure~\ref{fig:personalized-ad}, information about users' context can be utilized to create more captivating advertisements for them.

Further, the prediction module may interact with the modification module to commit to the modified output that mostly increases user's experience, \ie CTR and SR.
Even for SA, the prediction module may need to be made more sophisticated to predict the CTR and SR of the responsive ads.
This is in stark contrast to the standard DCO since it may not significantly change the ad image overall, but only slightly changes the part of it by replacing some ad assets.
Correspondingly, the advertiser may value the generated ad differently, \ie write different bid, with respect to the quality of the ad generated by LLM-based DCO.
Overall, a vast amount of studies should be done to clearly investigate a plausible framework for LLMA, even for SA, to adapt the LLM-based DCO.

\vspace{0.3cm}
\noindent {\color{red}Challenge:} {\bf Meeting system requirements.}
Obviously, running LLM-based DCO should not entail the failure of the system requirements discussed in Section~\ref{sec:require}.
For instance, in terms of latency, we observed that generating an image with the GPT4 takes almost about $10$s for Figure~\ref{fig:personalized-ad}.
Given that the online advertising typically requires the ad latency up to $100$ms, the required latency is almost $100$ times the conventional threshold.
Thus, the fundamental research challenge is to enable a fast production of an ad image to meet the conventional latency requirement in the online advertising system.
Further, it will be technically interesting to balance the tradeoff between the ad's quality and latency which governs the advertiser and user utility, respectively.

% Finally, we remark that these involved process demand additional computational resources, potentially impacting latency when incorporating responsive ads.
% Online advertisement typically requires the overall latency of real time bidding process to be within $100$ms, however, we experience several second of delay to modify the image output presented in Figure~\ref{fig:personalized-ad}. 
% Thus, it is technically interesting to balance the tradeoff between the ad's quality and latency, \eg naively by using cache, and also it might be possible to pass along part of such cost to the advertisers via monetary charge.

Interestingly, as acute readers may have noticed, Figure~\ref{fig:personalized-ad} includes some technical issues in the images themselves.
For example, the first image in the car ads generate a word that is not easily understandable.
Indeed, LLM-based DCO may induce the reliability issue due to the hallucinating nature of LLM \citep{Ji_2023, guerreiro2023hallucinations, bubeck2023sparks}.\footnote{\ssedit{One might view this as a simply inability of a model tailored to the text-to-image generation. We refer to~\cite{daras2022discovering} for a discussion on it.}}
Thus, LLM-based DCO should ensure that it does not generate a hallucinated ad that may hurt the advertiser satisfaction.
A simple approach might be to require LLM to minimally revise the reference images, requesting the advertisers to submit several images for reference.

% \krcomment{Keivan: write something about hallucination, LLM alignment, etc in temrs of research challenges. Cite some papers. ALso discuss Travel ad and point out some issues.
% }

% {\color{red}Challenge:} {\bf Reliability and LLM alignment.}
% }

% {\bf Research question: Challenges in modification, prediction, bidding.}

% \sscomment{Add one more para discussing issues in the figure. And then discuss research challenge.}

% \sscomment{todo: polish this para}
% \ssdelete{
% Beyond the modification module, both the prediction and bidding modules require adjustments, given the integration of user context in new ads.
% Bidders now face the task of not only assessing the quality of the modified output in terms of advertisements but also determining its appeal to the user.
% Moreover, as new modified outputs have the potential to be more enticing for users, CTR and SR may witness an increase.
% Consequently, the prediction modules also need to take user context into account and evaluate whether personalized ads can enhance CTR and SR or not.
% }

% In our framework, the feasibility of responsive ads is evident, albeit with the trade-off of increased complexity in the system.
% It's essential to acknowledge that 

\vspace{0.3cm}
\noindent {\color{red}Challenge:} {\bf Cost sharing model.}
Unlike from the traditional DCO technique that simply determines the efficient combination of ad assets, LLM-based responsive ad can do much beyond by creating truly new content.
On the other hand, such process necessarily entails increased use of computational resource, especially since more queries to LLM will be required to customize the ads.
We here discuss several plausible cost sharing models that LLMA might adopt.

The simplest one will be to charge the advertisers whenever the contents are dynamically modified.
For example, the advertiser and LLMA commit to a mutual contract on how much the advertiser pay the LLMA for each responsive ad, and LLMA charges the advertiser whenever it responsively modifies the ad contents.
As the advertiser might hope to use responsive ad only if the user context is sufficiently relevant to its ads, the contract might specify when the advertiser's ads will be responsively changed.
Also, LLMA may provide several options for advertiser to determine how frequently and largely it should responsively modify the ads.

This model would work well for those charged by cost-per-impression, however, advertisers with cost-per-click or cost-per-conversion may not willingly commit to the contract since they desire to pay the platform only if the actual click or conversion happen, even though the platform indeed responsively modified its ads for every impression.
LLMA might give several options to the advertisers under which event they will be charged to pay for responsive ad.
Ads with cost-per-click model may still want to pay for responsive ads per impression or per click.
For the latter, LLMA may offer a larger amount of commission per click by predicting the average CTR of the ads.

LLMA must also decide whether or not to account the increased amount of payment for each advertiser as the increment of the actual bid or just regard it as a separate commission.
For the former case, the advertisers who are willing to charge more for the responsive ad commission fee will take advantage of getting higher chance of delivering their ads, compared to those who do not use responsive ad option.
By doing so, the auction module can select the ad that truly maximize the total revenue expected from delivering a single advertisement, thereby increasing the platform's revenue.

\end{document}